\documentclass[]{article}
\input{Qcircuit}
\usepackage{booktabs,graphicx,rotating}
\newcommand{\s}[2]{\sigma_{#1}^{(#2)}} 
\newcommand{\hf}{\frac{1}{2}}
\newcommand{\tv}{\frac{3}{2}}
\newcommand{\Stot}{\ensuremath{S_{\mathrm{tot}}}}
\newcommand{\Sxtot}{\ensuremath{S_{x,\mathrm{tot}}}}
\newcommand{\Sytot}{\ensuremath{S_{y,\mathrm{tot}}}}
\newcommand{\Sztot}{\ensuremath{S_{z,\mathrm{tot}}}}
\newcommand{\mga}{\multigate{1}{\!\!1\!\!}}
\newcommand{\mgb}{\multigate{1}{\!\!\hf\!\!}}
\newcommand{\mgc}{\multigate{1}{\!\!-\hf\!\!}}
\newcommand{\ga}{\ghost{\!\!1\!\!}}
\newcommand{\gb}{\ghost{\!\!\hf\!\!}}
\newcommand{\gc}{\ghost{\!\!-\hf\!\!}}

\title{Universal Quantum Computation and Leakage Reduction in the
  3-Qubit Decoherence Free Subsystem} 
\author{Bryan H. Fong\\
HRL Laboratories, LLC\\
3011 Malibu Canyon Road, Malibu, CA 90265
\\ \\
Stephen M. Wandzura\\
California Creative Computational Physics\\
27441 Freetown Lane, Agoura Hills, CA 91301
}

\begin{document}
\maketitle
\begin{abstract}
  We describe exchange-only universal quantum computation and leakage
  reduction in the 3-qubit decoherence free subsystem (DFS).  We
  discuss the angular momentum structure of the DFS, the proper forms
  for the DFS CNOT and leakage reduction operators in the total
  angular momentum basis, and new exchange-only pulse sequences for
  the CNOT and leakage reduction operators.  Our new DFS CNOT sequence
  requires 22 pulses in 13 time steps.  The DFS leakage reduction
  sequence, the first explicit leakage reduction sequence of its
  kind, requires 30 pulses in 20 time steps.  Although the search for
  sequences was performed numerically using a genetic algorithm, the
  solutions presented here are exact, with closed-form expressions.
\end{abstract}

\section{Introduction}
Interest and research in semiconductor quantum dots for quantum
information processing has continued to grow since the original
proposal by Loss and DiVincenzo \cite{Loss:1998ee,Hanson:2007os}.  In
semiconductor quantum dot systems, with a single electron spin as a
qubit, two qubit gates using electron exchange interactions can be
performed on sub-nanosecond time scales \cite{Petta:2005vn}.  In
contrast, single qubit operations based on electron spin resonance may
be two orders of magnitude slower or more
\cite{Koppens:2006vn,Nowack:2007ys}.  The slow and technically
challenging single electron operations can be avoided by using an
encoding in which exchange interactions give encoded universality.
The possibility of universal computation using only the exchange
interaction, and its connection to decoherence free (DF) subsystems,
has been demonstrated in \cite{Bacon:2000qf,Kempe:2001a,Kempe:2001b}.
Bacon et al.\ \cite{Bacon:2000qf} describe the 4-qubit DF subspace,
giving exchange-based Hamiltonians that generate encoded universal
computation.  Kempe et al.\ \cite{Kempe:2001a,Kempe:2001b} give a
general theory for decoherence free subspaces and subsystems, and
prove the universality properties of these encodings employing only
exchange or other two-body interactions.  Experiments on a single
three-spin exchange-only encoded qubit have also recently been
reported \cite{Laird:2010}.

Any implementation of exchange-only quantum computing, however,
requires explicit gate sequences for a universal set of encoded gates.
For the 3-qubit decoherence free subspace, DiVincenzo et al.\
\cite{DiVincenzo:2000} found explicit exchange gate pulse sequences
for single encoded qubit gates, as well as for an encoded two-qubit
gate locally equivalent to a CNOT.  (A two-qubit gate is locally
equivalent to a second two-qubit gate if they differ only by single
qubit gates \cite{Makhlin:2002ve}.)  Kawano et al.\
\cite{Kawano:2005qf} showed that DiVincenzo's numerically obtained
locally equivalent CNOT sequence using 19 gate pulses in 13 time steps
in fact approximates a true, analytical, solution.

The DiVincenzo 19 gate locally equivalent CNOT pulse sequence is valid
only for the 3-qubit DF \emph{subspace}, and not for the entire
3-qubit DF \emph{subsystem}.  Computation in the DF subsystem offers
two major advantages compared to the subspace: initialization of
encoded subsystem states requires no magnetic field, and the subsystem
states are immune to all global decoherence mechanisms, not just
global decoherence in a single direction.  To exploit these features
of the 3-qubit subsystem, an analytic CNOT gate sequence for the
entire subsystem was found by Bonesteel et al. \cite{Bonesteel:2010},
which requires approximately 50 exchange gates.  Using a genetic
algorithm, we have significantly improved on the Bonesteel et
al. solution, finding a new analytic pulse sequence for the encoded
subsystem CNOT that requires just 22 exchange gates in 13 time steps.
Additionally, our solution is for the full encoded CNOT, and not just
the locally equivalent CNOT.  The 4-qubit DF subspace possesses the
same magnetic field-free initialization and global decoherence
immunity as the 3-qubit subsystem.  An encoded CNOT gate sequence has
also been found for the 4-qubit encoding \cite{Hsieh:2003}, requiring
50 exchange gates in 27 time steps.  With a CNOT gate sequence that is
more than twice as expensive as our 3-qubit CNOT solution, and the
additional overhead of an extra physical qubit per encoded qubit, the
4-qubit encoding is apparently inferior to the 3-qubit encoding.

Computation with the 3-qubit DF subsystem requires not only a
universal set of encoded gates, but also an effective means of
recovering from errors.  Though the subsystem protects against global
decoherence, local decoherence mechanisms such as nuclear hyperfine
and electron-electron dipole-dipole coupling still exist, and give
rise to both encoded errors and leakage from the encoded subsystem.
Encoded errors can be corrected using standard quantum error
correction procedures \cite{Preskill:1998jw}, but leakage errors must
be converted, or reduced, to encoded errors in the course of applying
quantum error correction.  The incorporation of leakage reduction
units into fault tolerant quantum error correction circuits is
described in \cite{Aliferis:2007bh}.  Kempe et al.\ \cite{Kempe:2001b}
describe a ``SWAP If Leaked'' (SIL) operator that performs leakage
reduction for the 3-qubit DF subspace with only exchange gates: given
a potentially leaked encoded qubit $A$ and a fiducial, unleaked
encoded qubit $B$, the SIL operator leaves $A$ unchanged if it is
unleaked, and replaces $A$ with a valid encoded state if it is leaked.
Kempe et al.\ discuss the action of an SIL operator on encoded basis
states but do not give an explicit pulse sequence for SIL.  Here we
present the first exchange-only pulse sequence for leakage reduction
in the 3-qubit DF subsystem, requiring 30 exchange gates in 20 time
steps, with closed-form pulse timings.  Together, our leakage
reduction and CNOT pulse sequences provide two of the required
operations for computation using the 3-qubit DFS.

The paper is organized as follows.  In Section 2 we review the angular
momentum structure and basis states of one and two 3-qubit DF
subsystems and define the notation to be used in the following
sections.  In Section 3 we give the form for the encoded CNOT operator
in the total angular momentum basis and the exchange gate sequence
satisfying this form.  Section 4 gives the analogous form and sequence
for the leakage reduction operator.  Section 5 describes the genetic
algorithm used in the numerical searches for the pulse sequences.  We
conclude in Section 6.  In the remainder of the paper we use the
acronym ``DFS'' to denote decoherence free subsystem.

\section{Angular Momentum Structure of DFS}
\label{sec:angular_momentum}
The basis states of the Hilbert space containing three qubits can be
described by three angular momentum quantum numbers $S$, $S_{1,2}$,
and $S_z$ \cite{Kempe:2001a}.  $S$ is the total spin of the three
qubits, corresponding to the eigenvalues of the operator $\mathbf{S}^2
\equiv \mathbf{S}_x^2 + \mathbf{S}_y^2 + \mathbf{S}_z^2$, where
$\mathbf{S}_j \equiv 1/2(\s{j}{1} + \s{j}{2} + \s{j}{3})$ is the total
angular momentum in the $j$ direction, $\s{j}{n}$ is the
$j^\mathrm{th}$ Pauli matrix on the $n^\mathrm{th}$ qubit, and $\hbar\equiv1$.
The eigenvalues of $\mathbf{S}^2$ are $S(S+1)$.  $S_{1,2}$ is
similarly the total spin of the first two qubits, and $S_z$ the total
$z$ spin of all three qubits. The eigenstates described by
these quantum numbers are shown in Table \ref{table:DFS1}.
\begin{table}
\[
\begin{array}[]{l*{8}{r}}
\toprule
         & 1   & 2   & 3   & 4   & 5   & 6   & 7   & 8\\
\midrule
S      & \hf & \hf & \hf & \hf & \tv & \tv & \tv & \tv \\
S_{1,2} & 0   & 0   & 1   & 1   & 1   & 1   & 1   & 1 \\
S_z      &\hf & -\hf &\hf & -\hf &\tv &\hf & -\hf &-\tv \\
\bottomrule
\end{array}
\]
\caption{Quantum numbers of commuting operators that uniquely specify
  all basis vectors in a three qubit, eight-dimensional Hilbert space.  $S$ is
  the total spin of all three qubits and specifies whether the DFS
  qubit has leaked; $S=\hf$ is unleaked, $S=\tv$ is leaked. $S_{1,2}$
  is the total spin of the first two qubits and gives the logical or encoded
  state for unleaked states.  $S_z$ is the total spin-$z$ of all three
  qubits and is the gauge quantum number.  The top line gives the
  index labels of the basis vectors, which correspond to
  Eqs.~(\ref{eq:ket1})--(\ref{eq:ket8}).}
\label{table:DFS1}
\end{table}
Their corresponding eigenvectors can be written in the computational
basis via Clebsch-Gordan coefficients:
\begin{eqnarray}
\ket{1}=&\frac{1}{\sqrt{2}}(\ket{010}-\ket{100})&=\ket{S_0}\ket{0}\label{eq:ket1}\\
\ket{2}=&\frac{1}{\sqrt{2}}(\ket{011}-\ket{101})&=\ket{S_0}\ket{1}\\
\ket{3}=&\sqrt{\frac{2}{3}}\ket{001}-\frac{1}{\sqrt{6}}\ket{010}
  -\frac{1}{\sqrt{6}}\ket{100}&=\frac{1}{\sqrt{3}}(\sqrt{2}\ket{T_+}\ket{1}-\ket{T_0}\ket{0})\\
\ket{4}=&\frac{1}{\sqrt{6}}\ket{011}+\frac{1}{\sqrt{6}}\ket{101}-\sqrt{\frac{2}{3}}\ket{110}&=
  \frac{1}{\sqrt{3}}(\ket{T_0}\ket{1}-\sqrt{2}\ket{T_-}\ket{0}\\
\ket{5}=&\ket{000}&=\ket{T_+}\ket{0}\\
\ket{6}=&\frac{1}{\sqrt{3}}(\ket{001}+\ket{010}+\ket{100})&=
  \frac{1}{\sqrt{3}}(\ket{T_+}\ket{1}+\sqrt{2}\ket{T_0}\ket{0})\\
\ket{7}=&\frac{1}{\sqrt{3}}(\ket{011}+\ket{101}+\ket{110})&=
  \frac{1}{\sqrt{3}}(\sqrt{2}\ket{T_0}\ket{1}+\ket{T_-}\ket{0})\\
\ket{8}=&\ket{111}&=\ket{T_-}\ket{1}\label{eq:ket8}
\end{eqnarray}
where the singlet $\ket{S_0}$ and triplet $\ket{T_\mu}$ states are defined
as:
\begin{eqnarray}
  \ket{S_0}&=&\frac{1}{\sqrt{2}}(\ket{01}-\ket{10})\\
\ket{T_+}&=&\ket{00}\\
\ket{T_0}&=&\frac{1}{\sqrt{2}}(\ket{01}+\ket{10})\\
\ket{T_-}&=&\ket{11}
\end{eqnarray}
The singlet and triplets are eigenstates of the $\mathbf{S}_{1,2}^2$
operator.

The 3-qubit DFS is spanned by the first four eigenstates in
Table~\ref{table:DFS1}.  The total spin quantum number $S$
distinguishes between states in the valid unleaked subspace ($S=\hf$)
and leaked states ($S=\tv$).  Within the unleaked subspace a valid
state of the 3-qubit DF subsystem has encoded (or logical) quantum
number $S_{1,2}$ and gauge quantum number $S_z$ that are unentangled:
valid states are factorizable states of the abstract (rather than
physical) subsystems corresponding to $S_{1,2}$ and $S_z$ quantum
numbers.  Valid DFS subsystem states are given by
$\alpha(\gamma\ket{1}+\delta\ket{2})+\beta(\gamma\ket{3}+\delta\ket{4})$.
In terms of the $S_{1,2}$ and $S_z$ quantum numbers we see the
factorizability explicitly:
\begin{eqnarray}
  \lefteqn{\alpha(\gamma\ket{1}+\delta\ket{2}) +
    \beta(\gamma\ket{3}+\delta\ket{4})}\nonumber\\   
  &=&\alpha(\gamma\ket{S=1/2,S_{1,2}=0,S_z=1/2}+\delta\ket{1/2,0,-1/2})
\nonumber\\
&&+\beta(\gamma\ket{1/2,1,1/2}+\delta\ket{1/2,1,-1/2})
\end{eqnarray}
Global decoherence mechanisms couple only to the $S_z$ quantum number,
and thus modify the gauge state only and not the encoded information.
A 3-qubit DFS state can be initialized as a singlet in the first two
qubits and an arbitrary spin in the third qubit, giving the state
$\gamma\ket{1}+\delta\ket{2}$.  The initialized state has encoded
quantum number $S_{1,2}=0$ and an undefined gauge state unentangled
with the encoded quantum number. Assuming that the 3-qubit DFS is not
leaked, measurement of the total spin $S_{1,2}$ on the first two
qubits distinguishes between singlet and triplet states and gives the
logical content of the DFS.

Encoded operations on the 3-qubit DFS are performed using the
exchange interaction between its constituent qubits.  The exchange
interaction between qubits $m$ and $n$ is defined as
$H^\mathrm{ex}_{m,n} \equiv 1/4(\s{x}{m} \cdot \s{x}{n} + \s{y}{m}\cdot\s{y}{n} +
\s{z}{m}\cdot\s{z}{n})$.  The exchange interaction generates the SWAP
operation between qubits $m$ and $n$, with a partial swap operation
given by
\begin{equation}
  U^\mathrm{ex}_{m,n}(p) =\exp(-i \pi p H^\mathrm{ex}_{m,n});
\label{eq:exchange_unitary}
\end{equation}
$U^\mathrm{ex}_{m,n}(p=\pm 1)$ gives the usual SWAP operation up to a global
phase.  Because the exchange interaction generates swaps between
qubits making up the 3-qubit DFS, it cannot change any total angular
momentum quantum numbers: it commutes with both $S$ and $S_z$ (as well
as $S_x$ and $S_y$).  The exchange interaction does change the
$S_{1,2}$ quantum number and hence the encoded state: exchange between
qubits 1 and 2 generates an encoded $z$ rotation, and exchange between
qubits 2 and 3 generates an encoded rotation about the
$\hat{n}=\{\sqrt{3}/2,0,-1/2\}$ axis.

The fact that the exchange interaction cannot change total angular
momentum quantum numbers motivates the choice for the set of quantum
numbers describing two 3-qubit DFS's or six physical qubits.  For DFS
qubits $A$ and $B$, a valid basis set is simply the ``product basis''
of the $A$ and $B$ eigenvectors given in Table~\ref{table:DFS1}.  None
of the quantum numbers of the product basis commutes with all of the
possible exchange operations between the six physical qubits.
However, if we use the total angular momentum basis consisting of
quantum numbers $\Stot$, $\Sztot$, $S_A$, $S_B$, $S_{A,1,2}$, and
$S_{B,1,2}$, the conservation of $\Stot$ and $\Sztot$ leads to a
partial diagonalization of the exchange-constructed operator and the
block structure described below.  $\Stot$ is the total spin of all six
physical qubits and $\Sztot$ is the total $z$ spin of all six physical
qubits.  $S_A$ ($S_B$) is the total spin of DFS qubit $A$ ($B$), with
$S_A=\hf$ and $S_B=\hf$ corresponding to unleaked states.  $S_{A,1,2}$
($S_{B,1,2}$) is the spin of the first two qubits of DFS qubit $A$
($B$) and gives the logical information encoded in DFS qubit $A$
($B$).

The structure of a six-qubit system is given in \cite{Kempe:2001b}.
It consists of five spin-0 subspaces, nine spin-1 subspaces, five
spin-2 subspaces, and one spin-3 subspace \cite{Mandel:1995xw}.  Each
total spin subspace is further divided into $\Sztot$ subspaces.
Tables~\ref{table:DFS20}--\ref{table:DFS22} show the quantum numbers
for the basis states of $\Stot=0,1,2$, respectively ($\Stot=3$ is not
needed in the following).  Table~\ref{table:DFS20} gives the
five-dimensional spin-0 subspace of the six physical qubit system.
Table~\ref{table:DFS21} gives the nine-dimensional $\Stot=1$,
$\Sztot=-1$ subspace; $\Stot=1$, $\Sztot=0$ or 1 have analogous
nine-dimensional subspaces.  Table~\ref{table:DFS22} gives the
five-dimensional $\Stot=2$, $\Sztot=-2$ subspace; similarly, $\Stot=2$
and $\Sztot=-1$, 0, 1, or 2 have analogous five-dimensional subspaces.
The 64-dimensional basis vectors can again be written in terms of the
computational basis of six spin-$1/2$ particles using Clebsch-Gordan
coefficients.

Because the exchange interaction commutes with $\Stot$ and $\Sztot$,
any operator constructed from exchange gates cannot mix subspaces with
different $\Stot$ and $\Sztot$.  Additionally, because the exchange
interaction also commutes with $\Sxtot$ and $\Sytot$,
exchange-constructed operators must have exactly the same matrix
entries on subspaces with the same $\Stot$ but different $\Sztot$.
That is, the matrix entries of an exchange-constructed operator on the
$\Stot=1$, $\Sztot=-1$ subspace are exactly the same on the $\Stot=1$,
$\Sztot=0$ and $\Stot=1$, $\Sztot=1$ subspaces, and similarly for
$\Stot=2$, $\Sztot=-2$, -1, 0, 1, 2.  When written in the total
angular momentum basis, an exchange-constructed operator is a block
diagonal matrix, whose diagonal blocks are one 5$\times$5 spin-0
block, three identical 9$\times$9 spin-1 blocks, five identical
5$\times$5 spin-2 blocks, and a spin-3 block consisting of a
7$\times$7 identity matrix times a phase factor; none of these blocks
couples to another.  If we define the generator of swap to be
$H^\mathrm{sw}_{m,n}\equiv H^\mathrm{ex}_{m,n}-1/4$, so that $\exp(-I
\pi H^\mathrm{sw}_{m,n})$ gives a full SWAP, \textit{including} the
global phase, the swap-generated group is $SU(5) \times SU(9) \times
SU(5) \times U(1)$.  The $\Stot=3$ subspace transforms as the identity
under swaps, so the $U(1)$ phase factor must be $\exp(-i \theta [\Stot
(\Stot+1)-12])$.  The algebraic origin of the block structure is
described in \cite{Kempe:2001a,Kempe:2001b}.  Constructing DFS
operators on the two DFS qubits amounts to finding exchange gate
sequences that satisfy the desired forms of the block matrix in the
total angular momentum basis.

\begin{table}
\[
\begin{array}{lccccc}
\toprule
  & 1 & 2 & 3 & 4 & 5 \\
\midrule
 \Stot & 0 & 0 & 0 & 0 & 0 \\
 \Sztot & 0 & 0 & 0 & 0 & 0 \\
 S_A & \frac{1}{2} & \frac{1}{2} & \frac{1}{2} & \frac{1}{2} &
 \frac{3}{2} \\
\addlinespace[.02 true in]
 S_B & \frac{1}{2} & \frac{1}{2} & \frac{1}{2} & \frac{1}{2} & \frac{3}{2} \\
 S_{A,1,2} & 0 & 0 & 1 & 1 & 1 \\
 S_{B,1,2} & 0 & 1 & 0 & 1 & 1 \\
\bottomrule
\end{array}
\]
\caption{Quantum numbers for two DFS qubits in the total spin-0
  subspace.  $\Stot$ is the 
  total spin of the six physical qubits, 
  $\Sztot$ the total spin-$z$, $S_A$ ($S_B$) the total spin of DFS qubit
  $A$ ($B$), $S_{A,1,2}$ ($S_{B,1,2}$) the spin of the first two
  qubits of DFS qubit $A$ ($B$). The top row is the index label of the
  basis vectors in the total angular momentum basis.  Basis vectors
  1--4 are valid encoded states; basis vector 5 is a leaked state.}
\label{table:DFS20}
\end{table}

\begin{table}
\[
\begin{array}{lrrrrrrrrr}
\toprule
  & 6 & 7 & 8 & 9 & 10 & 11 & 12 & 13 & 14 \\
\midrule
 \Stot & 1 & 1 & 1 & 1 & 1 & 1 & 1 & 1 & 1 \\
 \Sztot & -1 & -1 & -1 & -1 & -1 & -1 & -1 & -1 & -1 \\
 S_A & \frac{1}{2} & \frac{1}{2} & \frac{1}{2} & \frac{1}{2} &
 \frac{1}{2} & \frac{1}{2} & \frac{3}{2} & \frac{3}{2} & \frac{3}{2}
 \\
\addlinespace[.02 true in]
 S_B & \frac{1}{2} & \frac{1}{2} & \frac{1}{2} & \frac{1}{2} & \frac{3}{2} & \frac{3}{2} & \frac{1}{2} & \frac{1}{2} & \frac{3}{2} \\
 S_{A,1,2} & 0 & 0 & 1 & 1 & 0 & 1 & 1 & 1 & 1 \\
 S_{B,1,2} & 0 & 1 & 0 & 1 & 1 & 1 & 0 & 1 & 1 \\
\bottomrule
\end{array}
\]
\caption{Quantum numbers for two DFS qubits in the total spin-1,
  $\Sztot=-1$  subspace.  Basis vectors 6--9 are valid encoded states; basis vectors
  are leaked states, with one or both of the constituent DFS
  qubits leaked.  Basis vectors
  10 and 11 are unleaked in DFS qubit $A$.
  See Table \ref{table:DFS20} for operator
  definitions.}
\label{table:DFS21}
\end{table}

\begin{table}
\[
\begin{array}{lrrrrr}
\toprule
  & 15 & 16 & 17 & 18 & 19 \\
\midrule
 \Stot & 2 & 2 & 2 & 2 & 2 \\
 \Sztot & -2 & -2 & -2 & -2 & -2 \\
 S_A & \frac{1}{2} & \frac{1}{2} & \frac{3}{2} & \frac{3}{2} & \frac{3}{2} \\
\addlinespace[.02 true in]
 S_B & \frac{3}{2} & \frac{3}{2} & \frac{1}{2} & \frac{1}{2} & \frac{3}{2} \\
 S_{A,1,2} & 0 & 1 & 1 & 1 & 1 \\
 S_{B,1,2} & 1 & 1 & 0 & 1 & 1 \\
\bottomrule
\end{array}
\]
\caption{Quantum numbers for two DFS qubits in the total spin-2,
  $\Sztot=-2$  subspace.  Basis vectors 15 and 16 are unleaked in DFS
  qubit $A$.  See Table \ref{table:DFS20} for operator
  definitions.}
\label{table:DFS22}
\end{table}

\section{DFS CNOT Pulse Sequence}
\label{sec:CNOT}
The CNOT operation on two unleaked 3-qubit DFS's must perform a CNOT
gate on the logical DFS information contained in the quantum numbers
$S_{A,1,2}$ and $S_{B,1,2}$, independent of the gauge state.  We seek
an exchange pulse sequence, i.e., a sequence of unitary operators
generated by the exchange interaction, whose product in the total
angular momentum basis gives the required CNOT operation on the
encoded quantum numbers.  Because of the structure of
exchange-constructed matrices described in 
Section~\ref{sec:angular_momentum}, we need only constrain the
$\Stot=0$, $\Sztot=0$, $5\times5$ block and $\Stot=1$, $\Sztot=-1$,
$9\times9$ block appropriately for the DFS CNOT.  In the $\Stot=0$,
$\Sztot=0$ block the CNOT matrix must take the following form:
\begin{equation}
e^{i \theta_0}
\left(
\begin{array}{cccc|c}
1 & 0 & 0 & 0 & 0 \\
0 & 1 & 0 & 0 & 0 \\
0 & 0 & 0 & 1 & 0 \\
0 & 0 & 1 & 0 & 0 \\
\hline
0 & 0 & 0 & 0 & e^{i\phi_0} 
\end{array}
\right),
\label{eq:CNOT0}
\end{equation}
where $\theta_0$ and $\phi_0$ are arbitrary phases, and the rows and
columns correspond to the basis vectors 1--5 in
Table~\ref{table:DFS20}.  The upper left $4\times4$ block is the usual
CNOT operation on the unleaked, logical states; unitarity requires
that the leaked state with $\Stot=0$, basis vector 5, be uncoupled
from the unleaked states.  In the three $9\times9$ spin-1 blocks the
CNOT matrix must be
\begin{equation}
e^{i \theta_1}
\left(
\begin{array}{cccc|ccccc}
1 & 0 & 0 & 0 & 0 & 0 & 0 & 0 & 0\\
0 & 1 & 0 & 0 & 0 & 0 & 0 & 0 & 0\\
0 & 0 & 0 & 1 & 0 & 0 & 0 & 0 & 0\\
0 & 0 & 1 & 0 & 0 & 0 & 0 & 0 & 0\\
\hline
0 & 0 & 0 & 0 &  c_{1,1} & c_{1,2} & c_{1,3} & c_{1,4} & c_{1,5} \\
0 & 0 & 0 & 0 &  c_{2,1} & c_{2,2} & c_{2,3} & c_{2,4} & c_{2,5} \\
0 & 0 & 0 & 0 &  c_{3,1} & c_{3,2} & c_{3,3} & c_{3,4} & c_{3,5} \\
0 & 0 & 0 & 0 &  c_{4,1} & c_{4,2} & c_{4,3} & c_{4,4} & c_{4,5} \\
0 & 0 & 0 & 0 &  c_{5,1} & c_{5,2} & c_{5,3} & c_{5,4} & c_{5,5} 
\end{array}
\right),
\label{eq:CNOT1}
\end{equation}
where again $\theta_1$ is an arbitrary phase and the rows and columns
correspond to basis vectors 6--14 in Table~\ref{table:DFS21}.  The
upper left $4\times4$ block must be the CNOT on the encoded basis
vectors, and the lower right $5\times5$ block $\{c_{i,j}\}$ is an
arbitrary unitary matrix on the leaked states 10--14.  Again, the
leaked and unleaked states must not couple.  The three $9\times9$
spin-1 blocks are automatically identical when constructed out of
exchange gates, so they need not be constrained separately.  The CNOT
matrix on the spin-2 and spin-3 subspaces is completely unconstrained,
aside from unitarity, which is automatically satisfied with the
exchange gates.

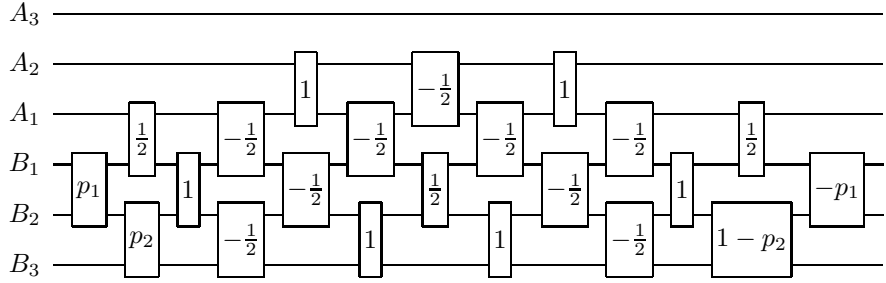
\begin{figure}
\hspace{2 em}\Qcircuit @C=.7em @R=1em @!R {
\lstick{A_3}  & \qw & \qw & \qw & \qw & \qw & \qw & \qw & \qw & \qw & \qw & \qw 
     & \qw & \qw & \qw \\
\lstick{A_2}  & \qw & \qw & \qw & \qw & \mga & \qw & \mgc & \qw & \mga & \qw & \qw
     &\qw &\qw & \qw\\
\lstick{A_1}  & \qw & \mgb & \qw & \mgc & \ga & \mgc & \gc & \mgc &
     \ga & \mgc & \qw  &\mgb &\qw & \qw\\
\lstick{B_1}  &\multigate{1}{\!\!p_1\!\!} & \gb & \mga & \gc & \mgc & \gc & \mgb & \gc &
      \mgc & \gc & \mga & \gb & \multigate{1}{\!\!-p_1\!\!} &\qw\\
\lstick{B_2}  &\ghost{\!\!p_1\!\!} & \multigate{1}{\!\!p_2\!\!} & \ga & \mgc & \gc & \mga & \gb & \mga
       & \gc & \mgc & \ga & \multigate{1}{\!\!1-p_2\!\!} & \ghost{\!\!-p_1\!\!} &\qw\\
\lstick{B_3}  & \qw & \ghost{\!\!p_2\!\!} & \qw & \gc & \qw & \ga & \qw & \ga & \qw & \gc &
  \qw  & \ghost{\!\!1-p_2\!\!} & \qw &\qw
}
\caption{Twenty-two pulse, 13 time step, exchange gate sequence for
  DFS CNOT.  DFS qubits $A$ (control) and $B$ (target) are arranged as
  shown.  Subscripts on $A$ and $B$ label the constituent physical
  qubits.  Each gate corresponds to an exchange unitary
  $U^\mathrm{ex}_{m,n}(p)$ [Eq.~(\ref{eq:exchange_unitary})], with the
  swap powers $p$ displayed explicitly in the gates.  $p=1$
  corresponds to a full SWAP operation, up to a global phase.
  $p_1=\arccos(-1/\sqrt{3})/\pi$ and $p_2=\arcsin(1/3)/\pi$.}
\label{fig:CNOT}
\end{figure}

An objective function for the CNOT search is constructed from the
constraints shown in Eqs.~(\ref{eq:CNOT0}) and (\ref{eq:CNOT1}).  Let
$U$ be the product of exchange gate unitaries in the total angular
momentum basis, ordered so that basis vectors 1--5 correspond to
Table~\ref{table:DFS20} and basis vectors 6--14 correspond to
Table~\ref{table:DFS21}.  The objective function for the CNOT search
is
\begin{equation}
f_\mathrm{CNOT}(U) = \sqrt{2
-\frac{1}{4}\lvert U_{1,1}+U_{2,2}+U_{3,4}+U_{4,3} \rvert
-\frac{1}{4}\lvert U_{6,6}+U_{7,7}+U_{8,9}+U_{9,8} \rvert
}.
\label{eq:CNOT_objective}
\end{equation}
Since $U$ is unitary by construction, its entries have modulus at most
1.  The objective function is zero only when $U_{1,1}, U_{2,2},
U_{3,4}$, and $U_{4,3}$ have modulus 1 and a common phase, and
$U_{6,6}, U_{7,7}, U_{8,9}$, and $U_{9,8}$ also have modulus 1 and a
(generally different) common phase.  This objective function is used
in the genetic algorithm described in Section~\ref{sec:genetic}.  The
two DFS qubits $A$ (control) and $B$ (target) are laid out in a linear
array, in the order $A_3, A_2, A_1, B_1, B_2, B_3$.  Only nearest
neighbor exchange gates are permitted.

Figure~\ref{fig:CNOT} shows the best solution found by the genetic
algorithm requiring 22 pulses in 13 time steps.  Though the genetic
algorithm uses approximate (finite precision) numbers, the final
solution found is analytic, with the exchange gate powers shown in the
figure.  Writing the exchange unitaries given in Figure~\ref{fig:CNOT}
in the total angular momentum basis and taking their product yields
the following matrix for the spin-0 subspace:
\begin{equation}
e^{i \theta_C}
\left(
\begin{array}{cccc|c}
1 & 0 & 0 & 0 & 0 \\
0 & 1 & 0 & 0 & 0 \\
0 & 0 & 0 & 1 & 0 \\
0 & 0 & 1 & 0 & 0 \\
\hline
0 & 0 & 0 & 0 & -1
\end{array}
\right),
\label{eq:CNOT0_solution}
\end{equation}
and for the spin-1 subspaces:
\begin{equation}
e^{i \theta_C}
\left(
\begin{array}{cccc|ccccc}
1 & 0 & 0 & 0 & 0 & 0 & 0 & 0 & 0\\
0 & 1 & 0 & 0 & 0 & 0 & 0 & 0 & 0\\
0 & 0 & 0 & 1 & 0 & 0 & 0 & 0 & 0\\
0 & 0 & 1 & 0 & 0 & 0 & 0 & 0 & 0\\
\hline
0 & 0 & 0 & 0 &  -\frac{11}{16} & -\frac{5\sqrt{3}}{16} & 0 & 0 & -\frac{\sqrt{15}}{8} \\
0 & 0 & 0 & 0 &  -\frac{5\sqrt{3}}{16} & -\frac{1}{16} & 0 & 0 & \frac{3\sqrt{5}}{8} \\
0 & 0 & 0 & 0 & 0 & 0 & 0 & 1 & 0 \\
0 & 0 & 0 & 0 & 0 & 0 & 1 & 0 & 0 \\
0 & 0 & 0 & 0 & -\frac{\sqrt{15}}{8} & \frac{3\sqrt{5}}{8} & 0 & 0 & -\frac{1}{4}
\end{array}
\right).
\label{eq:CNOT1_solution}
\end{equation}
The spin-0 and spin-1 subspaces have the same global phase.  The
matrices in Eqs.~(\ref{eq:CNOT0_solution}) and
(\ref{eq:CNOT1_solution}) clearly satisfy the constraints given in
Eqs.~(\ref{eq:CNOT0}) and (\ref{eq:CNOT1}).  The pulse sequence shown
in Figure~\ref{fig:CNOT} thus gives the full (not merely locally
equivalent) CNOT solution for the 3-qubit DF subsystem.  Removing the
$p_1$, $p_2$, $-p_1$, and $1-p_2$ gates results in a locally
equivalent CNOT solution of 18 pulses in 11 time steps for the DF
subsystem, which is shorter than the locally equivalent CNOT pulse
sequence for the DF subspace only \cite{DiVincenzo:2000}.

\section{DFS Leakage Reduction Pulse Sequence}
\label{sec:SIL}
Given a possibly leaked DFS qubit $A$ and a fiducial, unleaked DFS
qubit $B$, the DFS leakage reduction operator has the following
specification.  If $A$ is unleaked (a superposition of basis states
1--4 in Table~\ref{table:DFS1}), leakage reduction leaves the encoded
quantum number $S_{A,1,2}$ unchanged, but may possibly alter the gauge
quantum number $S_{A,z}$.  If $A$ is leaked, leakage reduction returns
$A$ to any state in the unleaked subspace.  The state of DFS qubit $B$
after leakage reduction is unconstrained, and will generally be
leaked.  Since we desire an exchange-only leakage reduction operator,
the second, fiducial DFS qubit $B$ is required.  A single, leaked DFS
qubit has total spin $S=\tv$, which cannot be converted to an unleaked
$S=\hf$ state by exchange operations on the single DFS qubit alone.

Though the leakage reduction operator returns leaked states to the
valid unleaked subspace, the final unleaked state is not necessarily a
valid subsystem state---the reset state will not generally be a
factorizable state of the encoded qubit and gauge qubit.  After
leakage reduction of a leaked state in DFS qubit $A$, the $A$
wavefunction will be
\begin{equation}
\lvert\psi_A\rangle = 
\alpha_I \lvert\psi_\lambda\rangle\lvert\phi_{g,I}\rangle 
+\alpha_X X_\lambda \lvert\psi_\lambda\rangle\lvert\phi_{g,X}\rangle 
+\alpha_Y Y_\lambda \lvert\psi_\lambda\rangle\lvert\phi_{g,Y}\rangle 
+\alpha_Z Z_\lambda \lvert\psi_\lambda\rangle\lvert\phi_{g,Z}\rangle,
\label{eq:reset_wavefunction}
\end{equation}
where $\alpha_\mu$ are complex amplitudes, $\lvert\psi_\lambda\rangle$
is the wavefunction of the encoded qubit, $\lvert\phi_{g,\mu}\rangle$
are wavefunctions of the gauge qubit, and $X_\lambda$, $Y_\lambda$,
and $Z_\lambda$ are single encoded qubit Pauli operators acting on the
encoded wavefunction alone.  The leakage reduced state thus generally
appears as a superposition of the correct encoded state with
Pauli-errors on the encoded state coupled to different gauge states.
Standard quantum error correction will return such a superposition
state in the unleaked subspace to a valid factorizable subsystem
state.  Quantum error correction procedures are constructed from DFS
encoded gates such as the DFS CNOT, Hadamard, and Pauli gates.  All
the DFS gates have been constructed to act independently of the gauge
state, so that quantum error correction on the encoded parts of
Eq.~(\ref{eq:reset_wavefunction}) proceeds as usual.  A projective
measurement in the course of the error correction procedure selects
one of the terms in Eq.~(\ref{eq:reset_wavefunction}), resulting in a
valid factorizable DFS state.  A similar argument holds in the case of
fully coherent error correction.

The leakage reduction operator in the total angular momentum basis
must have the following form.  In the spin-0 subspace the leakage
reduction matrix must be
\begin{equation}
\left(
\begin{array}{ccccc}
d_{1,1} & d_{1,2} & 0 & 0 & 0 \\
d_{2,1} & d_{2,2} & 0 & 0 & 0 \\
0 & 0 &            d_{1,1} & d_{1,2} & 0 \\
0 & 0 &            d_{2,1} & d_{2,2} & 0 \\
0 & 0 & 0 & 0 & e^{i \phi}
\end{array}
\right),
\label{eq:SIL0}
\end{equation}
where $\{d_{i,j}\}$ is an arbitrary $2\times2$ unitary matrix.  The
upper left $4\times4$ block is the outer product of the identity on
the encoded quantum number of DFS qubit $A$ and an arbitrary unitary
matrix on the encoded quantum number of DFS qubit $B$.  It ensures
that the encoded quantum number of DFS qubit $A$ is unchanged, with no
constraint on the unitary evolution of DFS qubit $B$'s encoded quantum
number, in the $\Stot=0$, $\Sztot=0$ gauge state.  The decoupling of
the fifth basis vector from the others ensures that no leakage from
DFS qubit $A$ occurs.

In the spin-1 subspaces the leakage reduction matrix must be
\begin{equation}
\left(
\begin{array}{ccccccccc}
e_{1,1} & e_{1,2} & 0 & 0            & 0 & 0 & f_{1,1} e_{1,3} & f_{1,2} e_{1,3} & 0\\
e_{2,1} & e_{2,2} & 0 & 0            & 0 & 0 & f_{1,1} e_{2,3} & f_{1,2} e_{2,3} & 0\\
0 & 0            & e_{1,1} & e_{1,2} & 0 & 0 & f_{2,1} e_{1,3} & f_{2,2} e_{1,3} & 0\\
0 & 0            & e_{2,1} & e_{2,2} & 0 & 0 & f_{2,1} e_{2,3} & f_{2,2} e_{2,3} & 0\\
e_{3,1} & e_{3,2} & 0       & 0      & 0 & 0 & f_{1,1} e_{3,3} & f_{1,2} e_{3,3} & 0\\ 
0 & 0            & e_{3,1} & e_{3,2} & 0 & 0 & f_{2,1} e_{3,3} & f_{2,2} e_{3,3} & 0\\
0 & 0 & 0 & 0 & g_{1,1} & g_{1,2} & 0 & 0 & g_{1,3} \\
0 & 0 & 0 & 0 & g_{2,1} & g_{2,2} & 0 & 0 & g_{2,3} \\
0 & 0 & 0 & 0 & g_{3,1} & g_{3,2} & 0 & 0 & g_{3,3}
\end{array}
\right),
\label{eq:SIL1}
\end{equation}
where $\{e_{i,j}\}$, $\{f_{i,j}\}$, and $\{g_{i,j}\}$ are all
arbitrary unitary matrices.  The ordering of matrix entries again
corresponds to the basis vector ordering in Table~\ref{table:DFS21}.
Column 1 of Eq.~(\ref{eq:SIL1}) gives the action of the leakage
reduction operator on basis vector 6 in Table~\ref{table:DFS21}.
Basis vector 6 can be brought to basis vectors 6,7, and 10, in any
complex linear combination: this action preserves the encoded 0
quantum number of DFS qubit $A$ and keeps $A$ unleaked, while allowing
DFS qubit $B$ to change its encoded state, or even leak.  Basis vector
7 (column 2) has a similar evolution as basis vector 6.  Columns 3 and
4 must be the same as columns 1 and 2, albeit with shifted entries;
this ensures the identity is applied on the encoded quantum number of
DFS qubit $A$.  The constraints on columns 1--4 and the constraints on
the spin-0 matrix together ensure that the identity is applied on the
encoded quantum number of DFS qubit $A$ regardless of gauge state,
assuming an unleaked DFS qubit $B$.

The remaining columns of Eq.~(\ref{eq:SIL1}) repair leaked states of
DFS qubit $A$ in the spin-1 subspaces.  Columns 7 and 8 (basis vectors
12 and 13) bring a leaked DFS qubit $A$ into basis vectors 6--11,
which are all unleaked in DFS qubit $A$, leaving no components in the
leaked basis vectors 12--14.  Unitarity of Eq.~(\ref{eq:SIL1}) forces
the entwining of the $\{e_{i,j}\}$ and $\{f_{i,j}\}$ matrices.
Unitarity again forces columns 5,6, and 9 to have 0 entries in their
first six components, leaving $\{g_{i,j}\}$ to be an arbitrary $3\times3$
unitary matrix.

Leaked states of DFS qubit $A$ also exist in the spin-2 subspaces and
must be reduced to unleaked states.  In the spin-2 subspace the
leakage reduction matrix must be
\begin{equation}
\left(
\begin{array}{ccccc}
0 & 0 & h_{1,1} & h_{1,2} & 0 \\
0 & 0 & h_{2,1} & h_{2,2} & 0 \\
k_{1,1} & k_{1,2} & 0 & 0 & k_{1,3} \\
k_{2,1} & k_{2,2} & 0 & 0 & k_{2,3} \\
k_{3,1} & k_{3,2} & 0 & 0 & k_{3,3} 
\end{array}
\right),
\label{eq:SIL2}
\end{equation}
where $\{h_{i,j}\}$ and $\{k_{i,j}\}$ are arbitrary unitary matrices.
The ordering of matrix entries in Eq.~(\ref{eq:SIL2}) corresponds to
the basis vector ordering in Table~\ref{table:DFS22}.  Columns 3 and 4
ensure that basis vectors 17 and 18, which are leaked in DFS qubit $A$
and unleaked in DFS qubit $B$, are brought to basis vectors 15 and 16,
which are unleaked in DFS qubit $A$.  We stress that here and in the
other spin subspaces, DFS qubit $B$ must be unleaked, or leakage
reduction of DFS qubit $A$ will fail.  The spin-3 subspace of the
leakage reduction operator acts only on basis vectors that have both
DFS qubits leaked and is unconstrained aside from unitarity.

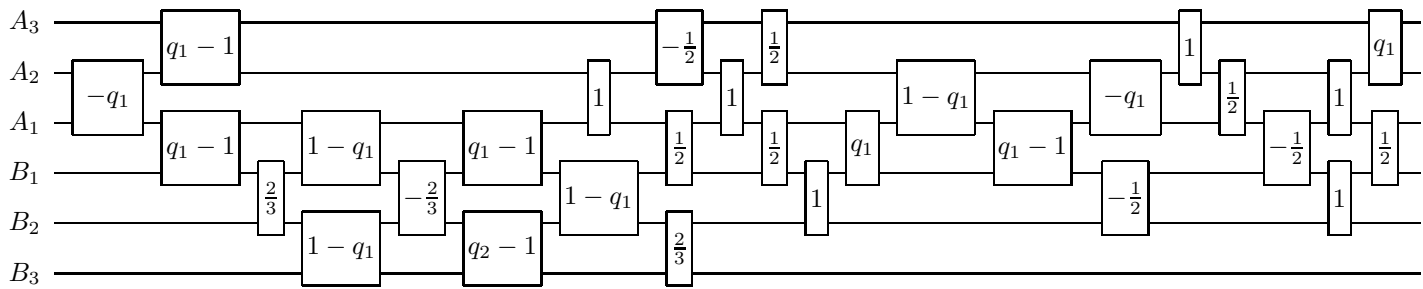
\begin{sidewaysfigure}
\hspace{2 em}\Qcircuit @C=.7em @R=1em @!R {
\lstick{A_3} & \qw  & \multigate{1}{\!\!q_1-1\!\!} & \qw  & \qw  & \qw  & \qw  & \qw  &
 \mgc & \qw  & \mgb & \qw  & \qw  & \qw  & \qw  & \qw  & \mga & \qw  &
 \qw  & \qw  & \multigate{1}{\!\!q_1\!\!} & \qw \\  
\lstick{A_2} & \multigate{1}{-q_1} & \ghost{\!\!q_1-1\!\!} & \qw  & \qw  & \qw  & \qw  &
  \mga & \gc & \mga & \gb & \qw  & \qw  & \multigate{1}{\!\!1-q_1\!\!} & \qw
  & \multigate{1}{-q_1} & \ga & \mgb & \qw  & \mga & \ghost{\!\!q_1\!\!} &
  \qw \\  
\lstick{A_1} & \ghost{-q_1} & \multigate{1}{\!\!q_1-1\!\!} & \qw  & \multigate{1}{\!\!1-q_1\!\!} &
 \qw  & \multigate{1}{\!\!q_1-1\!\!} & \ga & \mgb & \ga & \mgb & \qw  &
 \multigate{1}{\!\!q_1\!\!} & \ghost{\!\!1-q_1\!\!} & \multigate{1}{\!\!q_1-1\!\!} & 
 \ghost{-q_1} & \qw  & \gb & \mgc & \ga & \mgb & \qw \\ 
\lstick{B_1} & \qw  & \ghost{\!\!q_1-1\!\!} & \multigate{1}{\!\!\frac{2}{3}\!\!} & \ghost{\!\!1-q_1\!\!} &
 \multigate{1}{\!\!-\frac{2}{3}\!\!} & \ghost{\!\!q_1-1\!\!} & \multigate{1}{\!\!1-q_1\!\!} &
 \gb & \qw  & \gb & \mga & \ghost{\!\!q_1\!\!} & \qw  & \ghost{\!\!q_1-1\!\!} & \mgc &
 \qw  & \qw  & \gc & \mga & \gb & \qw\\ 
\lstick{B_2} & \qw  & \qw  & \ghost{\!\!\frac{2}{3}\!\!} & \multigate{1}{\!\!1-q_1\!\!} &
 \ghost{\!\!-\frac{2}{3}\!\!} & \multigate{1}{\!\!q_2-1\!\!} & \ghost{\!\!1-q_1\!\!} &
 \multigate{1}{\!\!\frac{2}{3}\!\!} & \qw  & \qw  & \ga & \qw  & \qw  & \qw  &
 \gc & \qw  & \qw  & \qw  & \ga  & \qw & \qw \\ 
\lstick{B_3} & \qw  & \qw  & \qw  & \ghost{\!\!1-q_1\!\!} & \qw  & \ghost{\!\!q_2-1\!\!} & \qw  &
 \ghost{\!\!\frac{2}{3}\!\!} & \qw & \qw  & \qw  & \qw  & \qw  & \qw  & \qw  &
 \qw  & \qw  & \qw  & \qw  & \qw & \qw
}
\caption{Thirty pulse, 20 time step, exchange gate sequence for DFS
  leakage reduction operator.  DFS qubits $A$ (potentially leaked) and
  $B$ (fiducial, unleaked) are arranged as shown.  Swap powers are
  displayed explicitly in the gates.  $q_1=\arccos(1/3)/\pi$ and
  $q_2=\arcsin(1/\sqrt{3})/\pi$.}
\label{fig:SIL}
\end{sidewaysfigure}

An objective function for the leakage reduction operator is
constructed from the constraints given in
Eqs.~(\ref{eq:SIL0})--(\ref{eq:SIL2}).  We additionally constrain
$\{f_{i,j}\}=\{h_{i,j}\}$, which forces a leaked state in DFS qubit
$A$ to be reset to a factorizable state of encoded and gauge quantum
numbers.  (This constraint is not necessary, but enabled us to find an
analytic solution.)  Again, let $U$ be the product of exchange gate
unitaries in the total angular momentum basis, and order the basis
vectors according to Tables~\ref{table:DFS20}--\ref{table:DFS22}.  We
define the following matrices, constructed from components of $U$:
\begin{eqnarray}
D^{(1)}&=&\left(
\begin{array}{cc}
U_{1,1} & U_{1,2} \\
U_{2,1} & U_{2,2}
\end{array}
\right),\\
D^{(2)}&=&\left(
\begin{array}{cc}
U_{3,3} & U_{3,4} \\
U_{4,3} & U_{4,4}
\end{array}
\right),\\
L^{(1)}&=&
\left(
\begin{array}{cc}
U_{6,12} & U_{6,13} \\
U_{8,12} & U_{8,13}
\end{array}
\right),\\
L^{(2)}&=&
\left(
\begin{array}{cc}
U_{7,12} & U_{7,13} \\
U_{9,12} & U_{9,13}
\end{array}
\right),\\
L^{(3)}&=&
\left(
\begin{array}{cc}
U_{10,12} & U_{10,13} \\
U_{11,12} & U_{11,13}
\end{array}
\right),\\
H&=&
\left(
\begin{array}{cc}
U_{15,17} & U_{15,18} \\
U_{16,17} & U_{16,18}
\end{array}
\right),
\end{eqnarray}
with $M^{(j)}=H^\dagger.L^{(j)}$, and
\begin{eqnarray}
E^{(1)}&=&\left(
\begin{array}{ccc}
U_{6,6} & U_{6,7} & M^{(1)}_{1,1}\\
U_{7,6} & U_{7,7} & M^{(2)}_{1,1}\\
U_{10,6} & U_{10,7} & M^{(3)}_{1,1}
\end{array}
\right),\\
E^{(2)}&=&\left(
\begin{array}{ccc}
U_{8,8} & U_{8,9} & M^{(1)}_{2,2} \\
U_{9,8} & U_{9,9} & M^{(2)}_{2,2}\\
U_{11,8} & U_{11,9} & M^{(3)}_{2,2}
\end{array}
\right).
\end{eqnarray}
With these definitions the leakage reduction operator objective
function is 
\begin{eqnarray}
f_\mathrm{LRO}(U) &=&
\left\|\frac{1}{4}(D^{(1)}+D^{(2)})^\dagger.(D^{(1)}+D^{(2)})-I_2
\right\|\nonumber\\
&&+\left\|\frac{1}{4}(E^{(1)}+E^{(2)})^\dagger.(E^{(1)}+E^{(2)})-I_3
\right\|\nonumber\\
&&+\left\|H^\dagger.H-I_2\right\|,
\label{eq:SIL_objective}
\end{eqnarray}
where $I_n$ is the $n\times n$ identity matrix.  The objective
function is zero when: $D^{(1)}=D^{(2)}$ and each is unitary,
satisfying Eq.~(\ref{eq:SIL0});
$H$ is unitary, satisfying Eq.~(\ref{eq:SIL2});
and $E^{(1)}=E^{(2)}$ and each is unitary, satisfying
Eq.~(\ref{eq:SIL1}) and the additional constraint that
$\{f_{i,j}\}=\{h_{i,j}\}$.

Figure~\ref{fig:SIL} shows the best solution found by the genetic
algorithm, requiring 30 pulses in 20 time steps.  Again, the solution
found is analytic, with the exchange gate powers shown in the figure.
In the total angular momentum basis the product of the exchange gate
unitaries satisfies the forms given in
Eqs.~(\ref{eq:SIL0})--(\ref{eq:SIL2}).  In Eq.~(\ref{eq:SIL0}) the
free phase satisfies $e^{i\phi}=1$, while the constituent unitary
matrices, up to a common global phase, are explicitly:
\begin{eqnarray}
\{d_{i,j}\} &=&
\left(
\begin{array}{cc}
 0 & \frac{1}{6} i \left(i+\sqrt{2}\right) \left(3 i+\sqrt{3}\right) \\
 (-1)^{5/6} & 0
\end{array}
\right),\\
\{e_{i,j}\} &=&
\left(
\begin{array}{ccc}
 0 & \frac{1}{6} i \left(i+\sqrt{2}\right) \left(3 i+\sqrt{3}\right) & 0 \\
 \frac{1}{6} \left(-i+\sqrt{3}\right) & 0 & -\frac{2 \sqrt{2}}{3} \\
 \frac{1}{3} \sqrt{2} \left(-i+\sqrt{3}\right) & 0 & \frac{1}{3}
\end{array}
\right),\\
\{f_{i,j}\} &=&
\left(
\begin{array}{cc}
 \frac{1}{12} \left(-i+2 \sqrt{2}\right) \left(3 i+\sqrt{3}\right) & \frac{1}{12} \left(1-i \sqrt{2}\right) \left(3
   i+\sqrt{3}\right) \\
 \frac{1}{4} \left(-i+\sqrt{3}\right) & -\frac{1}{4} \left(-i+\sqrt{2}\right) \left(-i+\sqrt{3}\right)
\end{array}
\right),\\
\{g_{i,j}\} &=&
\left(
\begin{array}{ccc}
 -\frac{1}{2} & \sqrt{-\frac{7}{12}+\frac{i \sqrt{2}}{3}} & 0 \\
 \sqrt{-\frac{7}{972}-\frac{i \sqrt{2}}{243}} & \frac{1}{18} & -\frac{4 \sqrt{5}}{9} \\
 -\sqrt{-\frac{140}{243}-\frac{80 i \sqrt{2}}{243}} & -\frac{2 \sqrt{5}}{9} & -\frac{1}{9}
\end{array}
\right),\\
\{h_{i,j}\} &=& \{f_{i,j}\},\\
\{k_{i,j}\} &=&
\left(
\begin{array}{ccc}
 \frac{1}{2} & -\sqrt{-\frac{7}{12}+\frac{i \sqrt{2}}{3}} & 0 \\
 -\sqrt{-\frac{7}{12}-\frac{i \sqrt{2}}{3}} & -\frac{1}{2} & 0 \\
 0 & 0 & -1
\end{array}
\right).
\end{eqnarray}

\section{Genetic Algorithm}
\label{sec:genetic}
The DFS CNOT and leakage reduction operator gate sequences were
constructed using the genetic method to minimze objective functions
associated with the desired CNOT and leakage reduction operator block
matrix forms.  Although simple genetic algorithms can be quite slow,
finding sequences of the lengths needed for the CNOT and leakage
reduction operators (fewer than 40 gates) can be accomplished using a
rather naive programming approach in a few weeks of evolution on an
Apple XServe computer.  (We also attempted optimization by simulated
annealing, but the gate sequences we present here for the CNOT and
leakage reduction operator were both found by genetic methods.) It
helps that the time consuming part of the computation is
embarrassingly parallel.

Our genetic approach iterates random changes on some population of
gate configurations, each given by a list of powers of swap between
adjacent physical qubits.  At each iteration (``generation'') we
augment the population with ``mutations'' and ``mating'', and then
apply ``natural selection'' that favors those configurations with low
value of the objective function.  We augment the objective functions
given in Eqs.~(\ref{eq:CNOT_objective}) and (\ref{eq:SIL_objective})
by adding a ``gate penalty'' term which is simply the length of the
sequence times some positive constant, the gate penalty parameter. The
gate penalty parameter is adjusted as the evolution progresses to give
a reasonable tradeoff between lowering the objective function and
lowering the gate count.

The code was written in {\em Mathematica} Version 7.0
\cite{Mathematica7:2008}, using the built-in parallel processing
capability. The two key {\em Mathematica} functions used were {\tt
  FindMinimum} and {\tt RandomSample}.  {\tt FindMinimum} searches for
a local minimum of a function using a variety of local descent
methods; {\tt RandomSample} gives a pseudorandom sampling of a list
with optionally weighted probability.  Further description of these
functions may be found in the {\em Mathematica} documentation
\cite{Mathematica7:2008}.

\subsection{Mutations}
We have programmed five different types of mutations, chosen at
random.  We have varied the probabilities of each mutation, but for
our CNOT and leakage reduction operator solutions, the probabilities
are as shown:
\begin{description}
\item[\tt RefineOne (8\%):] randomly chooses one gate of a
  configuration and uses \linebreak
  {\tt FindMinimum} to optimize its power of
  swap, starting with the current value.
\item[\tt RefineTwo (8\%):] randomly chooses an adjacent pair of gates
  and simultaneously minimizes the objective with respect to both
  powers of swap.
\item[\tt RefineAll (4\%):] runs {\tt FindMinimum} on all powers in the
  configuration. The time required to do this grows rapidly with the
  number of variables and is much larger than that of the previous two
  mutations. The {\em Mathematica} {\tt AccuracyGoal} and {\tt
    PrecisionGoal} {\tt Options} of {\tt FindMinimum} are set to
  modest values ($<5$) so that this operation runs in a tolerable
  length of time.
\item[\tt InsertGate (40\%):] inserts a gate at a random location with a
  random power, followed by {\tt RefineOne}.
\item[\tt DeleteGate (40\%):] deletes a gate at a random location,
  followed by {\tt RefineTwo} on the gates that were adjacent to the
  one removed. If the removed gate was the first or last, the {\tt
    RefineOne} is used.
\end{description}

\subsection{Mating}
The initial population (unmutated) is divided into pairs. We take the
beginning of one gate configuration and append the end of the other,
and then run {\tt RefineAll}. The length of the resultant
``offspring'' is chosen to be the minimum of that of the parent
configurations and the number of gates from each parent is chosen at
random. Because of {\tt RefineAll}, this is a time consuming
operation---but without the {\tt RefineAll} very few of the offspring
would survive natural selection. The genetic algorithm can be used
without mating, but the diversity of the population suffers.

\subsection{Natural Selection}
Surviving gate configurations are selected using the {\em Mathematica}
function \linebreak {\tt RandomSample}.  The sampling weight, which is
proportional to the probability of a member of the population
appearing in the sample, is given by the inverse of the objective
function augmented by the gate penalty term.  Configurations with a
small value of the objective function and low gate count are most
likely to survive.  The population size is specified by a parameter
under user control. After the survivors are determined, gates between
the same pair of qubits that are not separated by any noncommuting
gates are combined into one gate, by simply adding their powers.

\section{Conclusion}
Using a genetic algorithm, we have found new, exact exchange gate
pulse sequences for the encoded CNOT operation on two 3-qubit DFS
qubits, as well as for the DFS leakage reduction operator.  Our CNOT
solution gives the most efficient sequence yet found for either the
3-qubit subspace or subsystem, while our leakage reduction solution is
the first explicit sequence for any exchange-only subspace or
subsystem encoding.  We have found a wide range of other solutions for
the DFS CNOT and leakage reduction operator, but the sequences
presented here are the most efficient thus far.  Other solutions that
place additional constraints on the evolution of the leaked states may
have superior error propagation properties when combined with fault
tolerant error correction procedures; further investigation of such
sequences is underway.  

\section{Acknowledgements}
We thank Jim Harrington and Mark Gyure for their comments on the
manuscript.

Sponsored by United States Department of Defense.  The views and
conclusions contained in this document are those of the authors and
should not be interpreted as representing the official policies,
either expressly or implied, of the United States Department of
Defense or the U.S. Government.  Approved for public release,
distribution unlimited. 

\bibliographystyle{unsrt}

\end{document}